\documentclass[aps,prl,reprint,amsmath,amssymb,superscriptaddress, 10pt]{revtex4-2}
\usepackage[english]{babel}
\usepackage[dvipsnames]{xcolor}
\usepackage{amsmath,amssymb}
\usepackage{graphicx}
\usepackage{hyperref}
\usepackage{braket}
\usepackage{enumerate}
\usepackage{MnSymbol}
\usepackage{esint}
\usepackage{array}
\usepackage{lipsum}  
\usepackage{lineno}
\usepackage{mathrsfs}
\usepackage{siunitx}
\usepackage{booktabs}
\newcommand{\be}{\begin{equation}}
\newcommand{\ee}{\end{equation}}
\def\bes{\begin{subequations}}
\def\esu{\end{subequations}}

\definecolor{amber}{HTML}{CC9900}

\newcommand{\abstractA}{Microresonator frequency combs are key components for integrating optical devices into photonic circuits. They provide high stability, coherence, and low noise, even without external stabilization. Yet microcomb design remains largely heuristic: waveguide and resonator parameters are typically swept manually or semi-empirically, and the resulting spectra are evaluated only afterwards. This forward-design workflow is computationally costly, relies heavily on designer intuition, and does not generally identify optimal solutions. Here, we present an adjoint-based inverse-design framework for microresonator frequency combs that directly optimizes the comb spectrum with respect to pre-defined objectives. We demonstrate the power and flexibility of this approach by addressing three challenging problems: designing spectrally flat combs, synthesizing arbitrarily shaped comb spectra, and enforcing several performance metrics simultaneously through multi-objective optimization. Our results show that inverse design offers a systematic and efficient route to compact on-chip light sources with properties tailored to diverse applications. }

\begin{document}

\newcommand{\titleinfo}{Adjoint inverse design of microresonator frequency combs}
\title{\titleinfo}

%%%%%%%%%%%%%%%%%%%%%% List of authors %%%%%%%%%%%%%%%%%%%%%%
\author{Andrei Chuchalin}
 \affiliation{Laboratoire Temps-Fréquence, Université de Neuchâtel, Avenue de Bellevaux 51, Neuchâtel, Switzerland.}

 \author{Alexey Tikan}
  \email{alexey.tikan@unine.ch}
 \affiliation{Laboratoire Temps-Fréquence, Université de Neuchâtel, Avenue de Bellevaux 51, Neuchâtel, Switzerland.}

%%%%%%%%%%%%%%%%%%%%%%%%%%%%%%%%%%%%%%%%%%%%%%%%%%%%%%%%%%%%%%%%%%r

\begin{abstract}
\abstractA
\end{abstract}
\maketitle

Integrated photonic circuits (PICs) have become a central platform in the ongoing transformation of optical systems. By enabling compact, scalable, and energy-efficient functionality on chip, they support a broad range of applications, including high-capacity optical communications~\cite{hu2021ChipbasedOpticalFrequency}, neuromorphic and photonic computing~\cite{shastri2021PhotonicsArtificialIntelligence}, and emerging quantum technologies~\cite{moody2022RoadmapIntegrated}. Advances in foundry-compatible fabrication and wafer-scale integration further position PICs as a manufacturable technology platform capable of broadening access to high-performance optical devices~\cite{shekhar2024roadmapping,liu2021highyield}.

Microresonator frequency combs, also known as microcombs~\cite{herr2026FrequencyCombsCoherent}, have emerged as one of the key enabling technologies for PICs, providing a compact source of many mutually coherent wavelength channels and thereby supporting massively parallel operation in integrated optical systems~\cite{gaeta2019PhotonicChipBasedFrequency,sun2023ApplicationsOpticalMicrocombs}. Different applications, however, require different comb spectral profiles. In many parallel-channel PIC applications, a spectrally flat microcomb is highly desirable because it promotes uniform optical power per channel and helps maintain a comparable signal-to-noise ratio across channels~\cite{marinpalomo2017MicroresonatorBasedSolitons}. Another useful class of spectra is the sideband-enhanced comb, typically associated with dispersive-wave emission, also known as soliton Cherenkov radiation~\cite{akhmediev1995CherenkovRadiationEmitted}. Such spectra are particularly attractive for self-referencing and related frequency-metrology tasks, where strong comb lines at selected spectral locations can facilitate $f$--$2f$ or $2f$--$3f$ detection schemes~\cite{brasch2016PhotonicChipBased,yi2017SingleModeDispersive,brasch2017SelfReferenced}.

Various methods have been employed to control the spectral shape of microcombs, including coupled microresonators~\cite{tikan2021EmergentNonlinearPhenomena,tusnin2023NonlinearDynamicsKerr} and corrugated microresonators~\cite{yu2021SpontaneousPulseFormation}. The latter approach has enabled on-demand microcomb design by tailoring the dispersion profile on a mode-by-mode basis through photonic bandgap engineering~\cite{lucas2023TailoringMicrocombsInversedesigned}. In that work, direct simulations of the Lugiato-Lefever equation (LLE) were combined with a genetic algorithm to identify dispersion profiles that produce a desired microcomb spectrum.

Direct optimization of the microresonator dispersion in a mode-by-mode manner, for example through gradient-descent-based approaches, can be computationally expensive since accurate estimation of the resulting comb profile often requires accounting for hundreds or even thousands of modes. Fast and reliable minimization of an objective function that quantifies, for example, microcomb flatness, would therefore require evaluating the gradient of that function with respect to the displacement of each individual mode.

A similar computational challenge arises in shape-optimization problems in mechanics~\cite{allaire2001ShapeOptimizationHomogenization,bendsoe2004TopologyOptimization, thevenin_adjoint_2008}, hydrodynamics~\cite{mohammadi2009AppliedShapeOptimization,pironneau1974OptimumProfileStokes}, and photonics~\cite{molesky2018InverseDesignNanophotonics,lalaukeraly2013AdjointShapeOptimization, su_nanophotonic_2020, colburn_inverse_2021, dainese_shape_2024}. In photonics, these ideas have been applied directly to the optimization of PICs, whose constituent elements can be represented by high-dimensional parametrizations involving thousands of degrees of freedom~\cite{molesky2018InverseDesignNanophotonics,piggott2015InverseDesign,jensen2011TopologyOptimization}. A major advance in this area came with the introduction of adjoint-based inverse design~\cite{lalaukeraly2013AdjointShapeOptimization,piggott2015InverseDesign, su_nanophotonic_2020}, which enables efficient gradient evaluation with respect to all design parameters by solving only the direct and adjoint problems. As a result, the full gradient of the objective function can be obtained with only two simulations, up to a constant cost that is essentially independent of the number of design variables~\cite{lalaukeraly2013AdjointShapeOptimization,molesky2018InverseDesignNanophotonics,hughes2018AdjointMethodInverse}.

In this manuscript, we introduce an adjoint-based inverse-design framework for microcombs that enables efficient optimization of microresonator parameters with respect to pre-defined objectives. By combining direct and adjoint simulations associated with the LLE, the method provides gradient information for a large number of design variables. We pay special attention to  soliton-specific aspects of the optimization, including stability verification and pinning condition. To demonstrate the power of the method, we apply this framework to three challenging problems: the design of spectrally flat microcombs, the synthesis of arbitrarily shaped microcombs, and multi-objective optimization in which several target metrics are enforced simultaneously. Our results establish adjoint inverse design as a powerful and systematic route toward on-demand microcomb engineering.

\begin{figure*}[t!]
\centering
\includegraphics[width=0.95\textwidth]{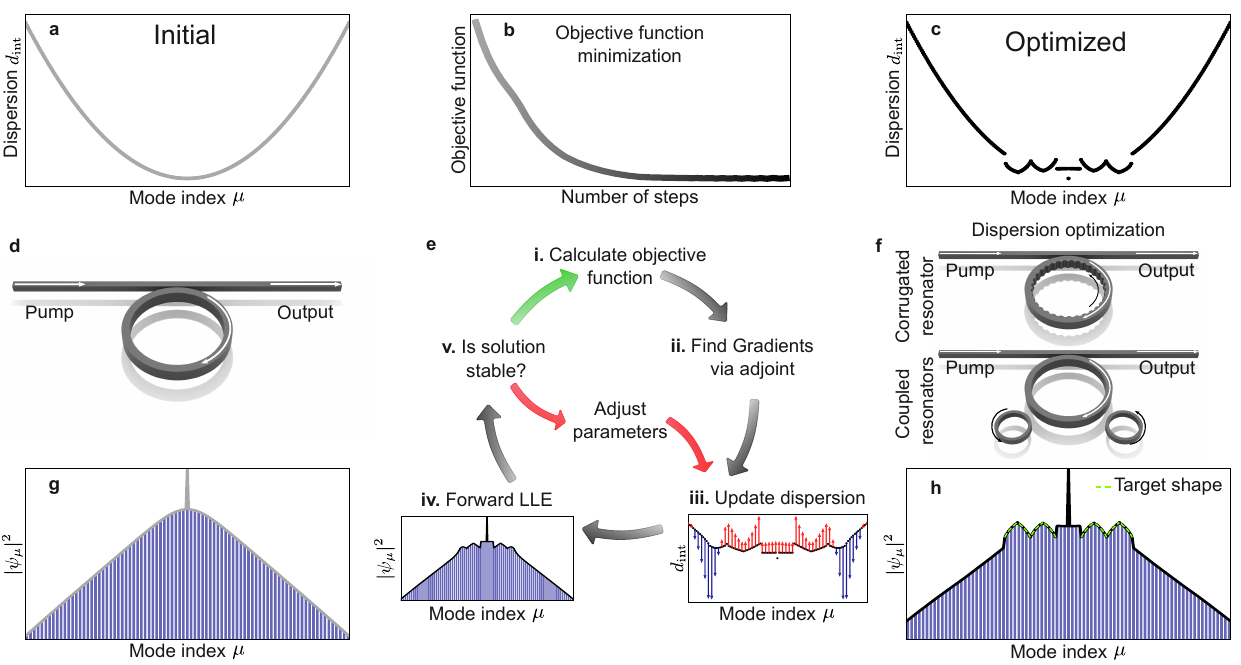}
\caption{\textbf{Optimization algorithm.}
\textbf{a,} The initial parabolic dispersion profile corresponding to the sech-type solution shown in \textbf{g}. \textbf{b,} Convergence of the objective function to its minima over optimization steps. \textbf{c,} Optimized integrated dispersion profile, corresponding to the comb target shape shown in \textbf{h}. \textbf{d,} Schematic of the platform with microring resonator having anomalous quadratic dispersion. \textbf{e,} The optimization loop consisting of five steps. \textbf{(i)} The objective function is evaluated for a current comb spectrum. \textbf{(ii)} Gradients of the objective function with respect to dispersion are computed efficiently using the adjoint method. \textbf{(iii)} The dispersion is updated via gradient descent method. \textbf{(iv)} The CME is solved for the updated dispersion and the comb spectrum is obtained. \textbf{(v)} The stability of the solution is checked, if it is found to be unstable, we adjust parameters and return back to the step \textbf{(iii)}. In case the solution is stable, we repeat the whole process until the loss function reaches its local minima. \textbf{f,} Schematic of two physical platforms capable to realize the optimized dispersion: a corrugated resonator (top) and a system of coupled resonators (bottom). \textbf{g,} Initial intracavity comb spectrum. \textbf{h,} Optimized intracavity comb spectrum. Green dashed line shows the target shape.
}
 \label{fig:algo_fig}
\end{figure*}

\section{Results}
\paragraph{\textbf{General setting for the adjoint problem.}} 
It is convenient to place the inverse design of microcombs to a more general context first. We can state that it implies minimizing an objective function subject to a complex-valued differential equation that constrains the system dynamics. Such problems can be addressed using the Lagrange multiplier formalism~\cite{giles2000IntroductionAdjointApproach,johnson_adjoint_methods_2021}. The modified objective function is
\begin{equation}
\begin{aligned}
    \mathscr{L}(x,x^*,\lambda,\lambda^*,p) =& L(x,x^*,p)\\-&\lambda \cdot g(x,x^*,p)-\lambda^*\cdot g^*(x,x^*,p),
    \label{eq:lagrangian}
    \end{aligned}
\end{equation}
where $g(x,x^*,p)$ is given by the LLE or its Fourier domain version called the coupled mode equations (CME), $ L(x,x^*,p)$ is the objective function to be optimized, $\lambda$ is the Lagrange multiplier, $x$ and $x^*$ - is the state determined by the model and its conjugated, $p$ - design (control) parameters, $\cdot$ defines the appropriate inner product.
The adjoint equations, detailed in the Methods section, are obtained by imposing stationary of the modified objective function and choosing $\lambda$ so that the gradient with respect to the parameters can be evaluated without explicitly differentiating the state.

Thus, to find the gradient of the objective function with respect to the design parameters, we need to compute 
\begin{equation}
    \frac{d L}{dp}=\frac{\partial L}{\partial p}-2\mathcal{Re}\left(\lambda^T \cdot \frac{\partial g}{\partial p}\right),
    \label{eq:grad}
\end{equation}
where $\mathcal{Re}$ denotes real part. It becomes clear that the Lagrange multipliers in Eq.~\ref{eq:lagrangian} play the role of the adjoint variables which are employed in the photonic inverse design frameworks~\cite{molesky2018InverseDesignNanophotonics}.

\paragraph{\textbf{Adjoint problem for the CME.}} 
One of the most computationally challenging problems in microcomb design is the mode-by-mode dispersion optimization that involves hundreds and thousands of modes. This requires writing constrained in the form of the CME and using the integrated dispersion \cite{herr2026FrequencyCombsCoherent} as design parameters.

Applying Eq.~\ref{eq:grad} to the normalized coupled-mode formulation, as detailed in the Methods, yields the gradient of the objective function with respect to the integrated-dispersion profile:
\begin{equation}
    \frac{dL}{d\text{d}_{\text{int}}(\mu)}=-4\eta\mathcal{Im}\left(\lambda_\mu \psi_\mu \right)
    \label{eq:lle_grad_mt}
\end{equation}
where $d_{\mathrm{int}}$ denotes the normalized integrated dispersion, $\mu$ is the relative mode index with $\mu=0$ corresponding to the pumped mode, $\psi_\mu$ is the normalized intracavity modal amplitude, and $\mathcal{Im}$ denotes the imaginary part. The parameter $\eta=\kappa_{\mathrm{ex}}/\kappa$ is the coupling efficiency, with $\kappa_{\mathrm{ex}}$ and $\kappa$ denoting the external coupling rate and the total cavity loss rate, respectively.

\paragraph{\textbf{Numerical algorithm.}}

Equation~\ref{eq:lle_grad_mt} provides the gradient of the objective function with respect to the normalized integrated-dispersion profile, enabling gradient-based optimization of the comb spectrum~\cite{herzog2010AlgorithmsPDEconstrainedOptimization}. Because the coupled-mode equations describe a nonlinear dissipative system, the optimization loop must also control convergence of the nonlinear solver and verify the dynamical stability of the stationary state.

The full procedure is summarized in Fig.~\ref{fig:algo_fig}. We typically initialize the algorithm from a conventional single-soliton state supported by a parabolic integrated-dispersion profile, $d_{\mathrm{int}}(\mu)=d_2\mu^2/2$. At each iteration, the forward problem is solved with an additional phase-pinning condition, which removes the continuous translational degeneracy of the soliton and improves convergence for asymmetric dispersion profiles. The resulting stationary field is used to evaluate the objective function. The corresponding adjoint problem is then solved to obtain the gradient in Eq.~\ref{eq:lle_grad_mt}. The dispersion profile is updated according to
\begin{equation}
    d_{\mathrm{int}}^{(n+1)}(\mu)
    =
    d_{\mathrm{int}}^{(n)}(\mu)
    -
    \alpha
    \frac{dL}{d \,d_{\mathrm{int}}(\mu)},
    \label{eq:dispersion_update}
\end{equation}
where $\alpha$ is the step size and $n$ denotes the optimization iteration. The forward problem is subsequently re-solved for the updated dispersion profile, using the solution from the previous iteration as the initial guess. This continuation strategy promotes convergence to the same solution branch and suppresses jumps to unrelated stationary states.

A stability check is then performed by computing the eigenvalues of the Jacobian obtained by linearizing the coupled-mode equations around the stationary solution~\cite{herr2026FrequencyCombsCoherent}. A solution is classified as linearly stable when all Jacobian eigenvalues have non-positive real parts, up to numerical tolerance. If an eigenvalue with positive real part is detected, the step size is reduced and a local search is performed in the $(\zeta_0,f^2)$ plane, where $\zeta_0$ is the normalized pump--cavity detuning and $f$ is the normalized pump amplitude. The search is initialized near the current operating point and continued until a stable stationary solution is recovered. The objective function is then evaluated for the stabilized solution. This forward--adjoint--update loop is repeated until the objective function reaches a local minimum or the gradient norm falls below a prescribed tolerance.

For several objectives used below, the required partial derivatives can be obtained analytically. In practice, however, we compute these derivatives using automatic differentiation, which provides an implementation-independent way to evaluate derivatives of user-defined objective functions. Automatic differentiation is the same computational principle used for backpropagation in machine learning and is available in standard scientific-computing packages, including \texttt{autograd}~\cite{autograd}.

\begin{figure*}[t!]
\centering
\includegraphics[width=0.95\textwidth]{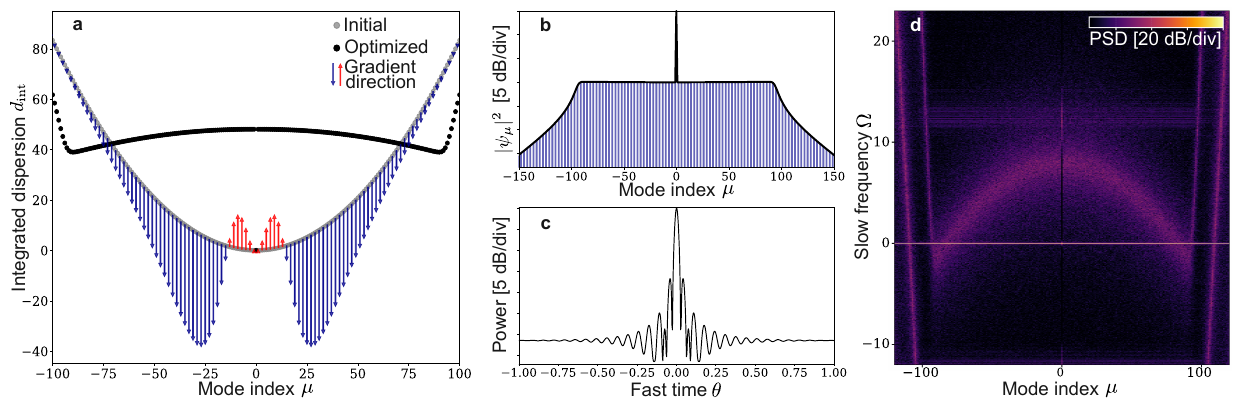}
\caption{ \textbf{Optimization of microcomb flatness.} \textbf{a,} Initial (gray dots) and optimized (black dots) integrated dispersion profiles as functions of mode index; gradients evaluated at the first iteration of the algorithm. \textbf{b,} Intracavity microcomb spectrum corresponding to the optimized dispersion, showing a flat-top region spanning approximately 200 modes. \textbf{c}, Resulting intracavity dissipative soliton state in the normalized fast time domain. \textbf{d,} Nonlinear dispersion relation, obtained by evolving the comb state in slow time and taking the Fourier transform.}
 \label{fig:flat_spectrum}
\end{figure*}

\paragraph{\textbf{Problem: flat-top microcomb.}}

Flat-top frequency combs are desirable in applications in which individual comb lines serve as parallel channels, including coherent optical communications~\cite{marinpalomo2017MicroresonatorBasedSolitons, hu2021ChipbasedOpticalFrequency}, LiDAR~\cite{Riemensberger2020}, and data center interconnects~\cite{rizzo2023MassivelyScalableKerr,rizzo2023PetabitScaleSiliconPhotonic,daudlin2025ThreedimensionalPhotonicIntegration,torrijos-moran2026IndustryInsightPhotonics}. In such systems, spectral flatness helps maintain a sufficiently high signal-to-noise ratio across the usable bandwidth. However, generating a microcomb with an approximately uniform line power over a prescribed spectral window remains a challenging design problem. In conventional forward design, flatness is typically assessed only after the resonator geometry and operating point have been chosen, making systematic optimization computationally inefficient.

Adjoint inverse design provides a direct route to this problem. Starting from a conventional Lugiato-Lefever equation soliton, we optimize the integrated-dispersion profile using the objective
\begin{equation}
    L_a=\sum_\mu \left(\frac{|f^{\text{out}}_\mu|^2}{P}-\frac{1}{N}\right)^2,
    \label{eq:obj_funct_flat}
\end{equation}
where  $f_\mu^{\text{out}}$ is the complex amplitude of the mode $\mu$ in the bus waveguide, $N$ is the number of modes considered in the simulation and $P$ is the total power of the frequency comb. Mode index $\mu$ changes from $-N/2$ to $N/2$. This objective function reaches its minimum value when all modes have equal power, thereby it controls the spectral flatness. 

Fig.~\ref{fig:flat_spectrum}a shows the initial integrated dispersion $d_\text{int}$ (gray dots) and the one obtained via adjoint inverse design method (black dots) over 90 iterations of the algorithm using 555 microresonator modes. The obtained solution was scaled to 2055 microresonator modes (see Methods). Red and blue arrows show the direction and the amplitude of gradient obtained by solving the adjoint Eq.~\ref{eq:lle_grad_mt} during the first iteration of the solver.

The resulting intracavity comb spectrum and the corresponding temporal waveform are shown in Fig.~\ref{fig:flat_spectrum}b,c. The optimized microcomb exhibits an approximately flat spectrum spanning about 200 modes. Consistent with Fourier duality, the corresponding temporal field resembles a sinc-like waveform on a finite continuous-wave background~\cite{xue2023DispersionlessKerrSolitons,turitsyn2020SolitonsincOpticalPulses}.

To identify the four-wave-mixing pathways and assess the dynamical stability beyond linear eigenvalue analysis of the stationary state, we compute the nonlinear dispersion relation (NDR)~\cite{leisman2019EffectiveDispersionFocusing,tikan2022NonlinearDispersionRelation,tikan2021EmergentNonlinearPhenomena,herr2012UniversalFormationDynamics}. The NDR, shown in Fig.~\ref{fig:flat_spectrum}d, is obtained as described in the Methods. The dissipative Kerr soliton appears as a horizontal line close to zero slow frequency. To make the weak linear-wave component visible, numerical noise is added at each step of the time-domain simulation; this reveals a dispersive branch that follows the optimized $d_{\mathrm{int}}(\mu)$ profile. Notably, although the optimized dispersion curve intersects the soliton branch near mode numbers $\mu \simeq \pm 100$, the resulting comb does not exhibit pronounced Cherenkov radiation\cite{brasch2016PhotonicChipBased}. This indicates that the inverse-designed dispersion modifies the phase-matching landscape of the four-wave-mixing process in a way that suppresses dispersive-wave emission over the optimized bandwidth keeping the microcomb profile flat to the $10^{-5}$  level precision.

\begin{figure*}[t!]
\centering
\includegraphics[width=0.95\textwidth]{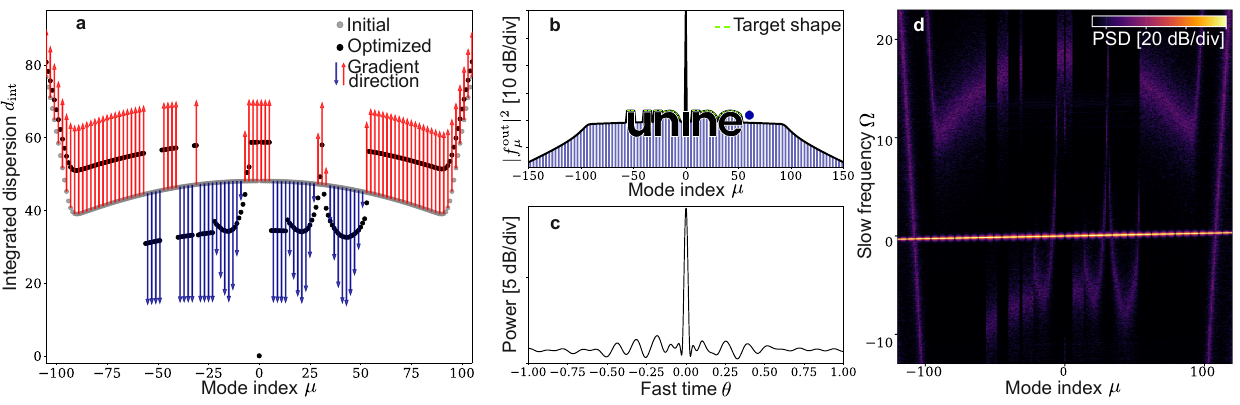}
\caption{ \textbf{Arbitrary shape optimization. a,} Initial (gray dots) and optimized (black dots) integrated dispersion profiles as functions of mode index; gradients are evaluated at the first iteration. \textbf{b,} Output optical spectrum corresponding to the optimized dispersion, showing the UNINE logo. Green dashed line shows the target shape. \textbf{c}, Resulting dissipative soliton state in the bus resonator in normalized fast time domain. \textbf{d,} Nonlinear dispersion relation, obtained by evolving the comb state in slow time and taking the Fourier transform. }
 \label{fig:unine_spectrum}
\end{figure*}

We note, the optimized flat comb is therefore not obtained from a flat integrated-dispersion profile. Instead, spectral flatness requires a non-trivial dispersion landscape that is identified by solving the inverse problem. Similar dispersion profiles may be accessible experimentally by shifting the pump mode in corrugated normal-dispersion microrings~\cite{yu2021SpontaneousPulseFormation} or by engineering mode hybridization in coupled microresonators~\cite{helgason2021DissipativeSolitonsPhotonic,Xue2015Normal-dispersionInteractions}. Recent investigations have shown the realization of nearly flat microcombs using these approaches~\cite{jin2026NanophotonicControlCollective,kolesnikova2026HybridizationKerrSolitons}.

\paragraph{\textbf{Problem: arbitrary comb shape.}} 
We next show that the same adjoint framework can synthesize a prescribed, non-uniform comb spectrum. Shaping the microcomb's spectrum is advantageous in a broad range of applications~\cite{Piciocchi:26}. In spectroscopy, spectral shaping allows tailoring the response function and simplifies the detection architecture in spectroscopic platforms~\cite{yoon_miniaturized_2022, deng_electrically_2022}. This method can also be applied to shape the soliton's temporal structure by considering an appropriate objective function, which is promising for coherent manipulations of quantum state excitations\cite{8843896,ma_precise_2020,nissila_driving_2021}.

As a representative target, we use the logo of the University of Neuch\^atel encoded as a modal power profile. This target is deliberately chosen because it contains both smooth regions and sharp spectral features, making it substantially more demanding than the flat-top spectrum considered above. The optimization is initialized from the flat comb state of Fig.~\ref{fig:flat_spectrum} to enhance the convergence and reduce the computation time.

The objective is defined pointwise over the target spectral window,
\begin{equation}
\label{eq:obj_2}
    L = \sum_\mu{(|f^\text{out}_\mu|^2-P_\mu)^2},
\end{equation}
where $P_\mu$ is the target power of mode $\mu$. This formulation allows controlling the power of each spectral component independently and engineer microcombs of an arbitrary shape.

Using the adjoint method optimization, we find the solution, in which the generated microcomb closely reproduces the target logo.
Figure~\ref{fig:unine_spectrum} shows the optimized integrated-dispersion profile together with the generated output spectrum, temporal profile, and the NDR. The inverse-designed comb reproduces the target modal pattern with high fidelity over the optimized bandwidth. The optimization converged over 20 iterations, using 555 microresonator modes. The obtained solution was scaled to 2055 microresonator modes (see Methods). The NDR suggests that the dissipative Kerr soliton acquired a group velocity which manifests itself as a tilt of the soliton line. This is expected in the asymmetric dispersion profiles.

We emphasize that the phrase ``arbitrary comb shape'' should be understood within the constraints of the nonlinear cavity dynamics. The target spectrum must be supported by the available bandwidth, the pump parameters, the cavity loss, and the physically realizable integrated-dispersion profile. Within these constraints, however, Eq.~\ref{eq:obj_2} provides a general objective for user-defined spectral-envelope synthesis. This capability extends the flat-comb design of Fig.~\ref{fig:flat_spectrum} from uniform power equalization to programmable spectral shaping of dissipative Kerr solitons.

\begin{figure*}[t!]
\centering
\includegraphics[width=0.95\textwidth]{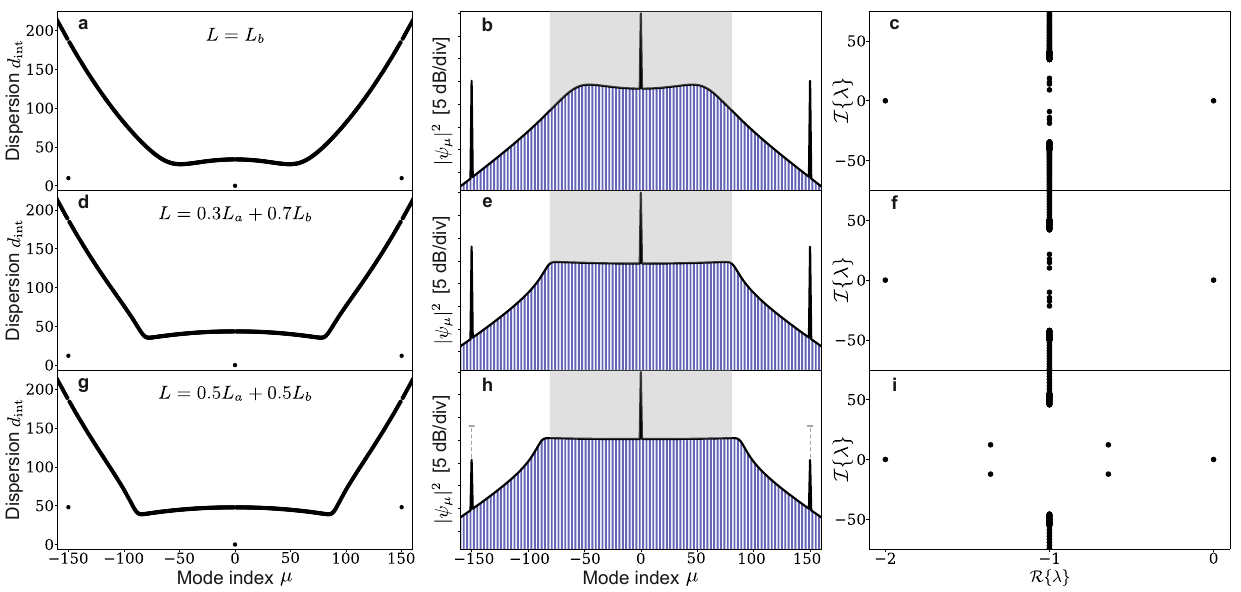}
\caption{ \textbf{Multi-objective optimization results. a,} Integrated dispersion profile for the transfer objective function $L_b$. \textbf{d,g} Integrated dispersion profiles for the combined objective functions, where $L_a$ (controlling flatness) and $L_b$ are taken with different weights. \textbf{b,e,h,} Spectra corresponding to the resulting dispersion profiles shown in \textbf{a,d,g} respectively. The gray line shows the expansion of flat-top region when the weight for $L_a$ is increased. In \textbf{h}, the dashed lines indicate the power of modes $\mu = \pm 150$ from panel \textbf{e} for direct comparison. \textbf{c,f,i,} Graphs showing eigenvalues of Jacobian corresponding to the obtained solutions. All of the eigenvalues have nonpositive real parts thus showing the stability of found integrated dispersions.}
 \label{fig:transfer}
\end{figure*}

\paragraph{\textbf{Problem: multi-objective optimization.}} 
The power of this method lies not only in its computational efficiency, but also in its versatility. The optimization demonstrated previously can be extended by adding another objective or another constraint. To illustrate this, we add to the objective function $L_a$ given by Eq.~\ref{eq:obj_funct_flat} controlling the flatness of the comb another one that controls the comb power enhancement at given mode numbers:
\begin{equation}
    L_b=  -\frac{|f^{\text{out}}_\mu|^2 +|f^{\text{out}}_{-\mu}|^2 }{P} 
    \label{eq:obj_funct_enh}
\end{equation}
We employ the objective function Eq.~\ref{eq:obj_funct_enh} to enhance sidebands that can be used for the microcomb stabilization via f-2f interferometry~\cite{cundiff2003ColloquiumFemtosecondOptical}.

Thus, the resulting objective function is given by
$L_\mathrm{tot} = aL_a  + bL_b $, where $a$ and $b$ are the weights. By changing weights, we can give preference to a certain objective. This combined function creates a unique combination of a flat comb in the middle and spectrally enhanced sidebands.  Fig.~\ref{fig:transfer} shows the optimization results for the combined objective functions $L_{\mathrm{tot}}$, where $L_{a}$ and $L_{b}$ are taken with different weights and pure transfer objective function $L_{b}$.

The objective function~\eqref{eq:obj_funct_enh} is minimized either when the numerator is maximized, leading to two peaks at $\mu=\pm150$ in power spectrum or when the denominator is minimized resulting in the emergence of almost flat region near the pump mode, with modes closer to pump having lower power. When the objective function $L_a$ controlling flatness is included with weights $a=0.3$ and $b=0.7$, we observe a remarkably enhanced flatness of central part of the microcomb, while maintaining the power transfer to $\mu = \pm 150$ modes. 
As the weight of the objective function $L_a$ is increased, the flat-top region is slightly expanded. However, since the $L_b$ objective function becomes less prioritized, the ratio of the power contained in the $\mu=\pm 150$ modes to the total comb power is decreased.

The multi-objective optimization presented here is general and versatile and it is not limited to utilizing the objective functions which are spectral in nature. For example, it is possible to consider objectives that are temporal (e.g., minimizing the pulse duration) or operational (e.g., penalizing integrated dispersion profiles, which are not smooth). Objective functions of different nature can be combined together via a weighted sum.

We further emphasize that the optimization problem is nonlinear and can admit multiple local optima. Therefore, the optimization outcome depends on the initial dispersion profile. In practice, the optimizer converges to a solution within the basin of attraction of the chosen initialization, rather than necessarily identifying the global optimum.

\section{Conclusion}

In conclusion, we have introduced an adjoint-based inverse-design framework for microresonator frequency combs that directly optimizes comb spectra according to user-defined objectives. By replacing heuristic parameter sweeps with gradient-based optimization, this approach enables microcomb design to be formulated as a systematic, objective-driven problem. We demonstrated the flexibility of the method through the design of spectrally flat combs, the synthesis of target spectral shapes, and the simultaneous optimization of multiple performance metrics.  These examples show that inverse design can provide both computational efficiency and design versatility in driven-dissipative nonlinear photonic systems. While this work focuses on dispersion optimization as one of the most computationally demanding problems, the adjoint method is also applicable to the optimization of other system parameters, including detuning, coupling coefficients, and cavity parameters.

More broadly, our results suggest that inverse design provides a general strategy for engineering driven-dissipative nonlinear photonic systems. By optimizing device parameters directly with respect to application-level objectives, this approach enables the design of compact on-chip light sources with tailored spectral properties. In this way, the framework offers an adaptable methodology for the inverse design of versatile nonlinear light sources, with potential relevance to a broad range of integrated photonic applications.

\section{Acknowledgements}
This work was supported by the Swiss National Science Foundation through SNSF Starting Grant (No.PS00P2\_234355).

\section{Methods}

\subsection{Adjoint method}
\label{sec:adj_method}
Let's consider a constrained optimization problem formulated using Lagrange multipliers. We suppose that the physical system is governed by the following equations:
\begin{equation}
    g_{i}(x,x^*,p)=0, \, i\in [1,N],
\end{equation}
where $x\in \mathbb{C}^N$ are the variables representing the physical state of the system (e.g., electric fields) and $p\in \mathbb{R}^P$ are the design parameters. The goal is to find such parameters $p$, which minimize the real-valued objective function $L(x,x^*,p)$. We introduce the Lagrange multipliers $\lambda \in \mathbb{C}^N$ and modify the objective function (a Lagrangian):
\begin{multline}
    \mathscr{L}(x,x^*,\lambda,\lambda^*,p) = L(x,x^*,p)-\\-\lambda \cdot g(x,x^*,p)-\lambda^*\cdot g^*(x,x^*,p)
\end{multline}
The derivative of $\mathscr{L}$ with respect to the design parameters thus becomes:
\begin{multline}
    \frac{d\mathscr{L}}{dp}=\frac{\partial L}{\partial p}-\lambda^T\frac{\partial g}{\partial p}-\lambda^{*T}\frac{\partial g^*}{\partial p}+\\+
    \left(\begin{pmatrix}
        \frac{\partial L}{\partial x} & \frac{\partial L}{\partial x^*}
    \end{pmatrix}-
    \begin{pmatrix}
        \lambda^T & \lambda^{*T}
    \end{pmatrix}\mathcal{J}\right)
    \begin{pmatrix}
        \frac{\partial x}{\partial p}\\\frac{\partial x^*}{\partial p}
    \end{pmatrix},
    \label{eq:lf_chain_rule}
\end{multline}
where $\mathcal{J}$ is the Jacobian matrix of $g(x,x^*,p)$. Eq.~\eqref{eq:lf_chain_rule} can be simplified if we find such Lagrange multipliers that:
\begin{equation}
\mathcal{J}
\begin{pmatrix}
    \lambda \\ \lambda^*
\end{pmatrix}=
\begin{pmatrix}
    \frac{\partial L^T}{\partial x}\\ \frac{\partial L^T}{\partial x^*}
\end{pmatrix}
\label{eq:adjoint_equation}
\end{equation}
Eq.~\eqref{eq:adjoint_equation} is called the adjoint equation. Once it is solved, the gradients become:
\begin{equation}
    \frac{d\mathscr{L}}{dp}=\frac{\partial L}{\partial p}-2\mathcal{Re}\left(\lambda^T\frac{\partial g}{\partial p}\right)
\end{equation}
The adjoint method is a powerful tool for solving gradient-based optimization problems, since it requires running only two full-field simulations regardless the number of design parameters $P$.

\subsection{\textbf{Derivation of the adjoint problem form CME}}
We utilize the adjoint method theory to solve the optimization problem, where waveguide dispersion acts as the design parameter and the loss function defines the target shape of the resulting frequency comb. The equations describing field inside the microresonator are the stationary coupled-mode equations (CME), expressed in normalized variables~\cite{herr2026FrequencyCombsCoherent}:
\begin{multline}
    g_\mu(\Psi(t),\Psi^*(t),d_\text{int}) = -(1+i(\zeta_0+d_{\text{int}}(\mu)))\psi_\mu+\\+i\mathscr{F}(|\Psi(t)|^2 \Psi(t))_\mu+f_0\delta_{\mu 0}, \, \mu \in [-N/2,N/2]
    \label{eq:cme}
\end{multline}
where $\Psi(t)$ is the slowly-varying electric field envelope, $\psi_\mu$ is its Fourier transform ($\mathscr{F}$), $\zeta_0$ is the normalized detuning, $d_{\text{int}}$ is the normalized integrated dispersion, $f_0$ is the normalized pump power and $N$ is the number of modes included in the simulation.

The output field becomes:
\begin{equation}
    f^{\text{out}}_{\mu} =f_0\delta_{\mu 0} - 2\eta \psi_{\mu},
\end{equation}
where $\eta$ is the coupling efficiency.

Since we optimize the frequency comb's shape, our objective function depends only on the output field, yielding:
\begin{equation}
    \frac{dL}{d\text{d}_\text{int}(\mu)}=-2\eta\left(\frac{\partial L}{\partial f_\mu^{\text{out}}}\frac{\partial \psi_\mu}{\partial \text{d}_\text{int}(\mu)}+\frac{\partial L}{\partial (f_\mu^{\text{out}})^*}\frac{\partial \psi_\mu^*}{\partial \text{d}_\text{int}(\mu)}\right)
\end{equation}
and the adjoint equation:
\begin{equation}
    \sum_\mu\left(\frac{\partial g_\mu}{\partial \psi_\nu}\lambda_\mu+\frac{\partial g^*_\mu}{\partial \psi_\nu}\lambda^*_\mu\right) = -\frac{\partial L}{\partial f_\nu^{\text{out}}}
\end{equation}
To handle the derivative of the nonlinear term, we apply the convolution theorem:
\begin{equation}
    \sum_\nu \lambda_\mu \frac{\partial}{\partial \psi_\nu}(|\Psi(t)|^2\Psi(t))_\mu=2\mathcal{F}(|\Psi(t)|^2\mathcal{F}^{-1}\{\lambda\})_{\nu}
\end{equation}
\begin{equation}
    \sum_\nu \lambda^*_\mu \frac{\partial}{\partial \psi_\nu}(|\Psi(t)|^2\Psi^*(t))_\mu=2\mathcal{F}((\Psi^*(t))^2\mathcal{F}^{-1}\{\lambda^*\})_{\nu}
\end{equation}
The adjoint equation becomes:
\begin{multline}
    -(1+i(\zeta_0+d_{\text{int}}(\nu)))\lambda_\nu+\\+i\mathcal{F}(2|\Psi(t)|^2\mathcal{F}^{-1}\{\lambda\}-(\Psi^*(t))^2\mathcal{F}^{-1}\{\lambda^*\})_\nu = \\=-\frac{\partial L}{\partial f^{\text{out}}_\nu}
\end{multline}
and the gradients become 
\begin{equation}
    \frac{dL}{d\text{d}_{\text{int}}(\mu)}=-4\eta\mathcal{Im}\left(\lambda_\mu \psi_\mu
    \right)
    \label{eq:lle_grad}
\end{equation}

\subsection{Simulation parameters}
For all simulations presented in the article, we used the following parameters: $D_1 = 100$~\si{GHz}, $D_2=1.5$~\si{MHz}, $\omega_0=2\pi\times192.2$~\si{THz}, $\kappa=300$~\si{MHz}, $n=2.4$, $n_2=2.4\cdot10^{-19}$~\si{m^2/W}, $\eta=0.5$, $V_\text{eff}=10^{-15}$~\si{m^3}, $\delta\omega =2\pi\times 1.05$~\si{GHz}, $P_{\text{in}}=0.7$~\si{W}, where $D_1$ is the free spectral range (FSR), $D_2$ is the second-order dispersion parameter, 
$\omega_0$ is the pump frequency, $\kappa$ is the total cavity loss rate, $n$ is the linear refractive index, $n_2$ is the Kerr nonlinear refractive index, $\eta$ is the coupling efficiency, $V_\text{eff}$ is the effective mode volume, $\delta \omega$ is the pump-cavity detuning and $P_\text{in}$ is the pump power. 

\subsection{Optimization algorithm}
\subsubsection{Forward method}
We perform a simple gradient descent algorithm to optimize the integrated dispersion to find the profile, which minimizes the loss function, hence producing the comb of a desired shape. First, we find the stationary solution of the coupled mode equations introduced in Eq.~\eqref{eq:cme}. We start the optimization from an initial integrated dispersion corresponding to anomalous group-velocity dispersion:
\begin{equation}
    d_{\text{int}}(\mu) = \frac{d_2}{2}\mu^2,
\end{equation}
where $d_{2}=\frac{2D_{2}}{\kappa}$. We consider the seed solution of the following form:
\begin{equation}
\Psi_{\text{seed}} = \mathcal{F}\left(\Psi_{\text{CW}}+\sqrt{2\zeta_0}e^{i\phi_0}\text{sech}\left(\sqrt{\frac{2\zeta_0}{d_2}}\phi\right)\right),
\end{equation}
where $\cos(\phi_0) = \frac{\sqrt{8\zeta_0}}{\pi f_{\text{0}}}$,
\begin{equation}
\Psi_{\text{CW}} = \frac{if_{\text{0}}}{\rho-\zeta_0+i}
\end{equation}
and $\rho$ corresponds to the lowest-branch solution of the cubic equation:
\begin{equation}
\rho^{3} - 2\zeta\rho^{2} +(\zeta_0^{2}+1)\rho-f_{\text{0}}^{2} = 0
\end{equation}
\subsubsection{Adjoint equation}
After the stationary solution is found, we proceed to solve the adjoint equation~\eqref{eq:adjoint_equation}. The Jacobian-vector product (JVP) is computed efficiently by using the SciPys's \texttt{LinearOperator} method avoiding storing the full Jacobian matrix. As a result, we get the gradients of the objective function with respect to the integrated dispersion.
\subsubsection{Gradient descent method}
After calculating the gradients~\eqref{eq:lle_grad}, we proceed with the gradient descent method, updating the disperison:
\begin{equation}
    d_{\mathrm{int}}^{(n+1)}(\mu)
    =
    d_{\mathrm{int}}^{(n)}(\mu)
    -
    \alpha
    \frac{dL}{d \,d_{\mathrm{int}}(\mu)},
    \label{eq:dispersion_update}
\end{equation}
where $\alpha$ is the step size, which can be chosen arbitrarily and $n$ denotes the optimization iteration. 

\subsubsection{Stability tracking}
After updating the dispersion, we repeat the whole process using the previous iteration solution as an initial guess (seed) for computing the updated one. 

However, since the forward method relies on the stationary solver, it is necessary to verify that the converged solution is stable. Thus, we added the additional step, which computes the eigenvalues of the Jacobian matrix and checks that all of them have nonpositive real parts. If the solution is found to be unstable, we return to the previous step and try to adjust the parameter $\alpha$ or to perform the local search in the $(\zeta_0, f^2)$ plane to restore stability. 

\subsection{Assymetric dispersion}

While the outlined algorithm works well for the symmetric dispersion, the asymmetric case gives rise to a non-zero group velocity. To account for this, we modify Eq.~\eqref{eq:cme} by introducing the additional term:
\begin{multline}
    g_\mu(\Psi(t),\Psi^*(t),d_\text{int},d_1) = -(1+i(\zeta_0+d_{\text{int}}(\mu)+d_1\mu))\psi_\mu+\\+i\mathscr{F}(|\Psi(t)|^2 \Psi(t))_\mu+f_0\delta_{\mu 0}, \, \mu \in [-N/2,N/2]
\end{multline}
and treat the normalized group velocity $d_1$ as another unknown. Now, the system consists of $N$ equations and $N+1$ unknowns. The additional equation can be chosen arbitrarily, but we use the pinning condition, which fixes the phase of the solution during the optimization. In time domain, it is written as:
\begin{equation}
    \frac{d}{d\alpha}\int_{-\pi}^{\pi}|\Psi(t+\alpha) - \Psi_{\text{seed}}(t)|^2 dt = 0
\end{equation} 
while in frequency domain it becomes:
\begin{equation}
    \mathcal{Im}\sum\mu \psi_\mu \psi^*_{\text{seed},\mu}=0
\end{equation}
Since we introduced the new variable, the adjoint equation is also modified:
\begin{multline}
        -(1+i(\zeta+d_{\text{int}}(\nu)+d_1\nu))\lambda_\nu+\\+i\mathcal{F}(2|\Psi(t)|^2\mathcal{F}^{-1}\{\lambda\}-(\Psi^*(t))^2\mathcal{F}^{-1}\{\lambda^*\})_\nu - \\ - \frac{i}{2}\nu \psi^*_{\text{seed},\nu}\tilde{\lambda} = -\frac{\partial L}{\partial f^{\text{out}}_\nu},
        \label{eq:adj_eq_mod}
\end{multline}
where $\tilde{\lambda}$ is the additional adjoint variable and the additional adjoint equation becomes:
\begin{equation}
    \sum_\mu \mu \mathcal{Im}(\lambda_\mu \psi_\mu) = 0
    \label{eq:additional_eq}
\end{equation}

\subsection{Application of automatic differentiation}

The right-hand side of adjoint equation~\eqref{eq:adjoint_equation} consists of the partial derivative of loss-function with respect to the output field. For certain loss functions, this derivative can be non-trivial to compute analytically. Therefore, we evaluate it using the \texttt{autograd} NumPy library, which automatically evaluates the required term once the loss function is specified.

\subsection{Nonlinear dispersion relation}
The nonlinear dispersion relation (NDR) can be computed by Fourier-transforming the $\psi_\mu(t)$. To reveal the resonance spectrum, we inject the white noise into Eq.~\eqref{eq:cme} and find the time evolution of each spectral component $\psi_\mu(t)$:
\begin{equation}
     \frac{d}{dt}\psi_{\mu}(t)=g_\mu(\Psi(t),\Psi^*(t),d_\text{int}) +\epsilon\xi_\mu(t),
     \label{eq:LLE_time_ev}
\end{equation}
where $\epsilon$ controls the noise amplitude and $\xi_\mu(t)$ is a complex-valued white noise process with independent real and imaginary parts distributed as $\mathcal{N}(0,1)$. Eq.~\eqref{eq:LLE_time_ev} is solved using the Runge-Kutta method of order 8. The injected noise gets enhanced at the effective resonance frequencies and the presence of coherent structures is revealed. We used $\epsilon=3\cdot 10^{-6}$ to plot NDR for Fig.~\ref{fig:flat_spectrum} and Fig.~\ref{fig:unine_spectrum}.
\subsection{Mode scaling}
Increasing spectral bandwidth leads to temporal compression of the corresponding soliton states. As a result, in order to resolve the soliton in time domain it is required to consider a large basis of modes.  

Although we can directly run the optimization algorithm for 2055 modes, we implemented a more computationally efficient algorithm to obtain the solution. The algorithm is based on the observation that gradients for the higher order modes $\mu$ are negligibly small, meaning that for these modes the integrated dispersion is the same with the initial anomalous quadratic dispersion. 

Therefore, we optimized the dispersion using an initial basis of 555 microresonator modes, subsequently extending the profile to 2055 modes by appending quadratic tails to the integrated dispersion. To construct a seed, the 555 mode stationary solution was transformed to the time domain, padded with continuous-wave background to the required resolution and then inverse-transformed back to the frequency domain. Finally, we used the stationary forward solver which converged to a solution following exactly the same shape with the 555 mode solution. 
\bibliography{biblio}
\end{document}